\documentclass{article}%
\usepackage{amsmath}
\usepackage{amsfonts}
\usepackage{amssymb}
\usepackage{graphicx}%
\providecommand{\U}[1]{\protect\rule{.1in}{.1in}}

\begin{document}

\author{Giuseppe Castagnoli
\and Pieve Ligure (Genoa), giuseppe.castagnoli@gmail.com}
\title{Quantum algorithms know in advance 50\% of the solution they will find in the
future }
\maketitle

\begin{abstract}
Quantum algorithms require less operations than classical algorithms. The
exact reason of this has not been pinpointed until now. Our explanation is
that quantum algorithms know in advance 50\% of the solution of the problem
they will find in the future. In fact they can be represented as the sum of
all the possible histories of a respective "advanced information classical
algorithm". This algorithm, given the advanced information (50\% of the bits
encoding the problem solution), performs the operations (oracle's queries)
still required to identify the solution. Each history corresponds to a
possible way of getting the advanced information and a possible result of
computing the missing information. This explanation of the quantum speed up
has an immediate practical consequence: the speed up comes from comparing two
classical algorithms, with and without advanced information, with no physics
involved. This simplification could open the way to a systematic exploration
of the possibilities of speed up.

\end{abstract}

\section{Introduction}

By integrating a set of notions developed in the series of articles $\left[
2\right]  $, $\left[  3\right]  $, $\left[  4\right]  $, and $\left[
5\right]  $, we provide a simple and self contained explanation of the quantum
speed up.

This should answer an existing need: Gross et al. assert that the exact
"reason" of the quantum speed up was never pinpointed, $\left[  9\right]  $,
(2009). Grover, with reference to a search in a database of size $N$, writes:
"What is the reason that one would expect that a quantum mechanical scheme
could accomplish the search in $\operatorname{O}\left(  \sqrt{N}\right)  $
steps? It would be insightful to have a simple two line argument for this
without having to describe the details of the search algorithm", $\left[
11\right]  ,$ (2001).

The explanation set forth in this article, in two lines, is: quantum
algorithms require a lower number of operations because they know in advance
50\% of the information about the solution of the problem they will find in
the future. The peculiar character of this explanation has to do with the
non-sequential behavior of the wave function, already highlighted by Dolev and
Elitzur in special interaction free measurement situations $\left[  8\right]
$. Here we show that this non-sequentiality is the crux of the quantum speed up.

In the sequel, quantum problem solving is seen as a game between two players:
the oracle and the quantum algorithm. The oracle chooses a function out of a
set of functions known to both players and gives to the second player the
black box for its computation. The second player should find out a certain
property of the function through function evaluation (oracle's query).

We show that a quantum algorithm: (i) requires the number of function
evaluations of a classical algorithm that knows in advance 50\% of the
information about solution of the problem and correspondingly (ii) can be
represented as the sum of all the possible histories of this classical
algorithm \ -- each history corresponds to a possible way of getting the
advanced information and a possible result of computing the missing
information. Thus the speed up comes from comparing two classical algorithms,
with and without advanced information. This brings the characterization of the
problems liable of being solved with a quantum speed up to an entirely
classical framework.

\section{Advanced knowledge}

We enter in media res by deriving point (i) of the introduction --\ from now
on called the \textit{50\% rule} --\ in a simple instance of Grover's data
base search algorithm $\left[  10\right]  $.

The oracle chooses a data base location -- an $n$\ bit string $\mathbf{k}%
\equiv\mathbf{~}k_{0},k_{1},...,k_{n-1}\in\left\{  0,1\right\}  ^{n}$\ (hides
a ball in drawer number $\mathbf{k}$) -- and gives to the second player the
black box that computes the Kronecker function $\delta\left(  \mathbf{k}%
,x\right)  $ (1 if $\mathbf{k}=x$, 0 otherwise). The second player has to find
the value of $\mathbf{k}$\ (the number of the drawer the ball is in) by
computing $\delta\left(  \mathbf{k},x\right)  $ for different values of $x$
(by opening different drawers). We also say: "by evaluating the function
$\delta\left(  \mathbf{k},x\right)  $".

The key step of our approach is representing together the production of the
problem on the part of the oracle and the production of the solution on the
part of the algorithm. We ideally add to the usual quantum registers $X$
(containing the argument of the function to query the black box with) and $V$
(hosting the result of function evaluation, $\operatorname{mod}2$ added to its
former content for logical reversibility) an auxiliary input register $K$
containing $\mathbf{k}$, the data base location chosen by the oracle. The
extended algorithm is: (0) prepare $K$ in the (even weighted)\ superposition
of all the values of $\mathbf{k}$, $X$ in the superposition of all the values
of $x$, and $V$ in the antisymmetric state, (1)\ perform function evaluation
and $\operatorname{mod}2$ add the result to the former content of $V$, and (2)
apply the transformation $U$ (see further below) to register $X$.

With $n=2$, the initial state is:%

\begin{equation}
\text{ }\Psi_{0}=\frac{1}{4\sqrt{2}}\left(  \left\vert 00\right\rangle
_{K}+\left\vert 01\right\rangle _{K}+\left\vert 10\right\rangle _{K}%
+\left\vert 11\right\rangle _{K}\right)  \left(  \left\vert 00\right\rangle
_{X}+\left\vert 01\right\rangle _{X}+\left\vert 10\right\rangle _{X}%
+\left\vert 11\right\rangle _{X}\right)  (\left\vert 0\right\rangle
_{V}-\left\vert 1\right\rangle _{V}). \label{preparation}%
\end{equation}
The superposition in $K$\ can indifferently be incoherent, in which case
$\left\vert 00\right\rangle _{K}$ should be replaced by $\operatorname{e}%
^{i\delta_{00}}\left\vert 00\right\rangle _{K}$, with $\delta_{00}$\ a random
variable with uniform distribution in $\left[  0,2\pi\right]  $, etc.

One function evaluation yields:%

\begin{equation}
\Psi_{1}=\frac{1}{4\sqrt{2}}\left[
\begin{array}
[c]{c}%
\left\vert 00\right\rangle _{K}\left(  -\left\vert 00\right\rangle
_{X}+\left\vert 01\right\rangle _{X}+\left\vert 10\right\rangle _{X}%
+\left\vert 11\right\rangle _{X}\right)  +\\
\left\vert 01\right\rangle _{K}\left(  \left\vert 00\right\rangle
_{X}-\left\vert 01\right\rangle _{X}+\left\vert 10\right\rangle _{X}%
+\left\vert 11\right\rangle _{X}\right)  +\\
\left\vert 10\right\rangle _{K}\left(  \left\vert 00\right\rangle
_{X}+\left\vert 01\right\rangle _{X}-\left\vert 10\right\rangle _{X}%
+\left\vert 11\right\rangle _{X}\right)  +\\
\left\vert 11\right\rangle _{K}\left(  \left\vert 00\right\rangle
_{X}+\left\vert 01\right\rangle _{X}+\left\vert 10\right\rangle _{X}%
-\left\vert 11\right\rangle _{X}\right)
\end{array}
\right]  (\left\vert 0\right\rangle _{V}-\left\vert 1\right\rangle _{V}),
\label{secondstage}%
\end{equation}
namely four orthogonal states of $K$\ , each a value of $\mathbf{k}$,
correlated with four orthogonal states of $X$, which means that the
information about the value of $\mathbf{k}$\ has propagated to register $X$. A
rotation of the\ measurement basis of $X$ makes this information readable. One
applies to $X$: \ (i) Hadamard transform, (ii)\ the transformation obtained by
computing $\delta\left(  0,x\right)  $, and (iii) another time Hadamard
transform (in the overall, the transformation $U$). This yields:%

\begin{equation}
\Psi_{2}=\frac{1}{2\sqrt{2}}\left(  \left\vert 00\right\rangle _{K}\left\vert
00\right\rangle _{X}+\left\vert 01\right\rangle _{K}\left\vert 01\right\rangle
_{X}+\left\vert 10\right\rangle _{K}\left\vert 10\right\rangle _{X}+\left\vert
11\right\rangle _{K}\left\vert 11\right\rangle _{X}\right)  (\left\vert
0\right\rangle _{V}-\left\vert 1\right\rangle _{V}), \label{final}%
\end{equation}
an entangled state where each value of $\mathbf{k}$ (in $K$) is correlated
with the corresponding solution found by the second player: the same value of
$\mathbf{k}$ in $X$.

We denote by $\left[  K\right]  $ and $\left[  X\right]  $\ the contents of
$K$ and\ $X$. Measurement of $\left[  K\right]  $ and $\left[  X\right]  $ in
(\ref{final}) determines the moves of both players: the oracle's choice (the
value of $\mathbf{k}$) and the solution provided by the second player. Note
that the state reduction\ induced by measuring $\left[  K\right]  $, backdated
to before running the algorithm, yields the original Grover's algorithm.

We discuss the behavior of the state vector. We see the nondeterministic
production of the contents of the two registers, due to measuring $\left[
K\right]  $ or $\left[  X\right]  $ in (\ref{final}), as \textit{mutual
determination} between such contents, like between two polarizations measured
in an entangled polarization state (mutual determination, or \textit{mutual
causality}, is between bits of information). The precise meaning of "mutual"
is specified by the following use of the term (see also $\left[  2\right]
,\left[  3\right]  ,\left[  4\right]  $):

We cannot say that reading the content of $K$ (i. e. the outcome of measuring
$\left[  K\right]  $)\ at the end of the algorithm causes the content of $X$
(the outcome of measuring $\left[  X\right]  $), namely that choosing the
drawer number (a value of $\mathbf{k}$) to hide the ball in on the part of the
oracle determines the drawer number the ball is found in by the second player
-- this is the classical perspective with no mutual determination.

For the same reason we cannot say that reading the content of $X$ at the end
of the algorithm causes the content of $K$, namely that reading the drawer
number at the end of the algorithm, on the part of the second player,
determines the drawer number chosen by the oracle, namely creates the ball in
the drawer with that number.

In consonance with time-symmetric quantum theory, we assume that mutual
causality is symmetrical $\left[  2\right]  $. Thus, the content of the two
registers is determined by reading the first (second) bit of register $K$\ and
the second (first) bit of register $X$. In this perspective, one bit of the
data base location is created by the oracle (by the action of measuring either
bit of $K$), the other bit by the action of reading, at the end of the
algorithm and on the part of the second player, the other bit of the data base
location in register $X$ (i. e. by the action of measuring the other bit of
$X$). It is important to notice that this other bit is the ball created in
that bit. Thus, the second player (the quantum algorithm) has to search only
the bit created by the oracle; this explains the fact that the quantum
algorithms requires just one function evaluation, namely the speed up of
Grover's algorithm for $n=2$.

Mutual causality is, in a different perspective, \textit{advanced knowledge}.
We should think of backdating to before running the algorithm the reduction
induced by measuring $\left[  K\right]  $. To the second player (to the
algorithm), this is indistinguishable from having a $\left[  K\right]  $
measured before running the algorithm -- see equations (\ref{preparation})
through (\ref{final}) -- thus to having a predetermined $\mathbf{k}$. In this
perspective the second player, by measuring $\left[  X\right]  $ at the end of
the algorithm, does not "create" any bit of information, he just "finds" the
two bits created by the oracle. Mutual causality becomes the quantum algorithm
knowing in advance, before running, 50\% of the information about the solution
it will produce at the end of the run.

In either form, mutual causality explains the structure of the quantum
algorithm. The quantum algorithm is an even weighted superposition of all the
possible histories of a classical algorithm that, knowing in advance 50\% of
the information about the solution of the problem, performs the function
evaluations still required to identify the solution. As clarified in the
sequel, each history corresponds to a possible way of getting the advanced
information and a possible result of computing the missing information.

We should make a specification. In Grover's algorithm, the outcomes of
measuring $\left[  K\right]  $ and $\left[  X\right]  $ in (\ref{final}) are
identical. The advanced information is indifferently 50\% of the content of
either register. This is not always the case in quantum algorithms, therefore
we define the advanced information, in relation to the measurement outcomes,
in a more general way. Since the content of $X$ is a function of the content
of $K$ (the solution is a function of the problem), the information contained
in $X$\ is redundant. Thus, the information encoding the solution of the
problem is all contained in register $K$, namely in the bit string
$\mathbf{k}$. Advanced knowledge concerns any two halves of $\mathbf{k}$.

In the following, for each one of the main quantum algorithms, we: (i) provide
the extended representation, (ii) pinpoint the half $\mathbf{k}$'s
representing advanced knowledge, (iii)\ check the 50\% rule, (iv)\ show that
the quantum algorithm can be represented as the sum of the histories of the
related advanced information classical algorithm, with history initial phases
chosen to reconstruct the quantum algorithm. We also try to rebuild the
quantum algorithm out of the advanced information classical algorithm using no
a priori knowledge of the quantum algorithm.

This latter part of the work contains a summary of $\left[  5\right]  $.
However, in $\left[  5\right]  $ we gave the 50\% rule mostly as a pattern
common to the main quantum algorithms. Now, algorithm by algorithm, we gear
the 50\% rule with the above explanation of the quantum speed up. Moreover, we
further develop the methodology for building the quantum algorithm out of the
advanced information classical algorithm.

\section{Deutsch's algorithm}

The set of functions is $f_{\mathbf{k}}:\left\{  0,1\right\}  \rightarrow
\left\{  0,1\right\}  $ -- table (\ref{table}).%

\begin{equation}%
\begin{tabular}
[c]{|c|c|c|c|c|}\hline
$x$ & $f_{00}(x)$ & $f_{01}(x)$ & $f_{10}(x)$ & $f_{11}(x)$\\\hline
0 & 0 & 0 & 1 & 1\\\hline
1 & 0 & 1 & 0 & 1\\\hline
\end{tabular}
, \label{table}%
\end{equation}
$\mathbf{k}\equiv k_{0},k_{1}$ is the table of the function (the sequence of
function values ordered for increasing values of the argument) and, clockwise
rotated, the suffix of the function. The oracle chooses at random one of these
functions and gives to the second player the black box that, given $x$,
computes $f_{\mathbf{k}}(x)$. The problem is finding whether the function is
"balanced" ($\mathbf{k}=01,10$) or constant. This requires two function
evaluations in the classical case, just one in the quantum case $\left[
6\right]  $.

We ideally add to the usual registers a two qubit register $K$ containing the
oracle's choice $\mathbf{k}$. Now the black box, given $\mathbf{k}$ and $x$,
computes $f\left(  \mathbf{k},x\right)  =f_{\mathbf{k}}(x)$.

The extended algorithm is: (0) prepare $K$ in the superposition of all
$\mathbf{k}$, $X$ in the superposition of all $x$, and $V$ in the
antisymmetric state, (1)\ perform function evaluation, and (2) apply Hadamard
to register $X$:\ %

\begin{equation}
\Psi_{0}=\frac{1}{4}\left(  \left\vert 00\right\rangle _{K}+\left\vert
01\right\rangle _{K}+\left\vert 10\right\rangle _{K}+\left\vert
11\right\rangle _{K}\right)  \left(  \left\vert 0\right\rangle _{X}+\left\vert
1\right\rangle _{X}\right)  \left(  \left\vert 0\right\rangle _{V}-\left\vert
1\right\rangle _{V}\right)  . \label{initial}%
\end{equation}
\ \ \ \ \ \ %

\begin{equation}
\Psi_{1}=\frac{1}{4}\left[  (\left\vert 00\right\rangle _{K}-\left\vert
11\right\rangle _{K})(\left\vert 0\right\rangle _{X}+\left\vert 1\right\rangle
_{X})+(\left\vert 01\right\rangle _{K}-\left\vert 10\right\rangle
_{K})(\left\vert 0\right\rangle _{X}-\left\vert 1\right\rangle _{X})\right]
\left(  \left\vert 0\right\rangle _{V}-\left\vert 1\right\rangle _{V}\right)
. \label{fin}%
\end{equation}

\begin{equation}
\Psi_{2}=\frac{1}{2\sqrt{2}}\left[  (\left\vert 00\right\rangle _{K}%
-\left\vert 11\right\rangle _{K})\left\vert 0\right\rangle _{X}+(\left\vert
01\right\rangle _{K}-\left\vert 10\right\rangle _{K})\left\vert 1\right\rangle
_{X}\right]  \left(  \left\vert 0\right\rangle _{V}-\left\vert 1\right\rangle
_{V}\right)  . \label{hadamard}%
\end{equation}

Measuring $\left[  K\right]  $ and $\left[  X\right]  $\ in (\ref{hadamard})
determines the moves of both players: the oracle's choice in register $K$ and
the solution found by the second player in register $X$: 0 if the function is
constant, 1 if balanced. Backdating to before running the algorithm the
reduction induced by measuring $\left[  K\right]  $\ gives the original
Deutsch's algorithm.

We check the 50\% rule. The information acquired by measuring $\left[
K\right]  $ in (\ref{hadamard})\ is the two bits read in register $K$ (the
content of register $X$ is redundant). There is advanced knowledge of anyone
of these two bits. The quantum algorithm requires the number of function
evaluations of a classical algorithm that knows in advance either
$k_{0}=f(\mathbf{k},0)$ or $k_{1}=f(\mathbf{k},1)$. To identify the character
of the function, this algorithm must acquire the other bit of information by
computing, respectively, either $k_{1}=f(\mathbf{k},1)$ or $k_{0}%
=f(\mathbf{k},0)$. Thus the advanced information classical algorithm has to
perform one function evaluation like the quantum algorithm, which verifies the
50\% rule.

This rule shows that Deutsch's problem is solvable with a quantum speed up
independently of our knowledge of the quantum algorithm. In fact the speed up
comes from comparing two classical algorithms, with and without the advanced
information. The same applies to all quantum algorithms.

We represent the quantum algorithm as an even weighted superposition of the
histories of the advanced information classical algorithm. Table (\ref{d})
gives the combinations of advanced information and result of computing the
missing information.%

\begin{equation}%
\begin{tabular}
[c]{|c|c|c|}\hline
\# & advanced information & result of function evaluation\\\hline
1 & $k_{0}=0$ & $k_{1}=f(\mathbf{k},1)=0$\\\hline
2 & $k_{0}=0$ & $k_{1}=f(\mathbf{k},1)=1$\\\hline
3 & $k_{0}=1$ & $k_{1}=f(\mathbf{k},1)=0$\\\hline
4 & $k_{0}=1$ & $k_{1}=f(\mathbf{k},1)=1$\\\hline
5 & $k_{1}=0$ & $k_{0}=f(\mathbf{k},0)=0$\\\hline
6 & $k_{1}=0$ & $k_{0}=f(\mathbf{k},0)=1$\\\hline
7 & $k_{1}=1$ & $k_{0}=f(\mathbf{k},0)=0$\\\hline
8 & $k_{1}=1$ & $k_{0}=f(\mathbf{k},0)=1$\\\hline
\end{tabular}
\label{d}%
\end{equation}

Each classical computation history is represented in quantum notation as a
sequence of sharp states:

\begin{itemize}
\item Row \#1. Advanced information $k_{0}=0$. The classical algorithm should
compute $k_{1}=f(\mathbf{k},1)$ that, for this row, is $k_{1}=0$. The quantum
representation of the oracle's choice is thus $\left\vert 00\right\rangle
_{K}$. To compute $f(\mathbf{k},1)$, the initial state of the input register
$X$ must be $\left\vert 1\right\rangle _{X}$. Since the result is
$\operatorname{mod}2$ added to the initial content of register $V$, we should
split row \#1 into: \#1.1 with $V$ initially in $\left\vert 0\right\rangle
_{V}$ and \#1.2 with $V$ initially in $\left\vert 1\right\rangle _{V}$. The
initial state of \#1.1 is $\Psi_{0}^{(1.1)}=\left\vert 00\right\rangle
_{K}\left\vert 1\right\rangle _{X}\left\vert 0\right\rangle _{V}$, that of
\#1.2 is $\Psi_{0}^{(1.2)}=-\left\vert 00\right\rangle _{K}\left\vert
1\right\rangle _{X}\left\vert 1\right\rangle _{V}$. Computation histories have
to be added and must be given an initial phase. We set the initial phases in
such a way that, in the superposition of all histories, we obtain the initial
state of the quantum algorithm. This show that the quantum algorithm can be
represented as an even weighted superposition of all the possible histories of
the advanced information classical algorithm.\ The sum of the initial states
of \#1.1 and \#1.2 is: $\Psi_{0}^{(1)}=\Psi_{0}^{(1.1)}+\Psi_{0}%
^{(1.2)}=\left\vert 00\right\rangle _{K}\left\vert 1\right\rangle _{X}\left(
\left\vert 0\right\rangle _{V}-\left\vert 1\right\rangle _{V}\right)  $. We
normalize at the end. Function evaluation transforms $\Psi_{0}^{(1)}$ into
itself: $\Psi_{1}^{(1)}=\Psi_{0}^{(1)}$($\operatorname{mod}2$ adding $f\left(
00,1\right)  =0$ to the former content of $V$ leaves this content unaltered).

\item Row \#5. Advanced information $k_{1}=0$, result of function evaluation
$k_{0}=0$. The same rationale yields: $\Psi_{0}^{(5)}=\left\vert
00\right\rangle _{K}\left\vert 0\right\rangle _{X}\left(  \left\vert
0\right\rangle _{V}-\left\vert 1\right\rangle _{V}\right)  $, $\Psi
_{1}^{\left(  5\right)  }=\Psi_{0}^{(5)}$.

\item The sum of \#1 and \#5 yields the transformation of $\left\vert
00\right\rangle _{K}\left(  \left\vert 0\right\rangle _{X}+\left\vert
1\right\rangle _{X}\right)  \left(  \left\vert 0\right\rangle _{V}-\left\vert
1\right\rangle _{V}\right)  $ into itself, i. e. the function evaluation stage
of Deutsch's algorithm when $K$ is in $\left\vert 00\right\rangle _{K}$.

\item Row \#2. $\Psi_{0}^{(2)}=\left\vert 01\right\rangle _{K}\left\vert
1\right\rangle _{X}\left(  \left\vert 0\right\rangle _{V}-\left\vert
1\right\rangle _{V}\right)  $, $\Psi_{1}^{(2)}=-\Psi_{0}^{(2)}$
($\operatorname{mod}2$ adding $f\left(  01,1\right)  =1$ to the former content
of $V$ swaps $\left\vert 0\right\rangle _{V}$ and $\left\vert 1\right\rangle
_{V}$ -- rotates by $\pi$ the phase of the pair of histories).

\item Row \#7. $\Psi_{0}^{(7)}=\left\vert 01\right\rangle _{K}\left\vert
0\right\rangle _{X}\left(  \left\vert 0\right\rangle _{V}-\left\vert
1\right\rangle _{V}\right)  $, $\Psi_{1}^{(7)}=\Psi_{0}^{(7)}$
($\operatorname{mod}2 $ adding $f\left(  01,0\right)  =0$ to the former
content of $V$ leaves this content unaltered).

\item The sum of \#2 and \#7 yields the transformation of $\left\vert
01\right\rangle _{K}\left(  \left\vert 0\right\rangle _{X}+\left\vert
1\right\rangle _{X}\right)  \left(  \left\vert 0\right\rangle _{V}-\left\vert
1\right\rangle _{V}\right)  $ into $\left\vert 01\right\rangle _{K}\left(
\left\vert 0\right\rangle _{X}-\left\vert 1\right\rangle _{X}\right)  \left(
\left\vert 0\right\rangle _{V}-\left\vert 1\right\rangle _{V}\right)  $,
namely the function evaluation stage of Deutsch's algorithm when $K$ is in
$\left\vert 01\right\rangle _{K}$.

\item Proceeding in a similar way with the other histories, summing over, and
normalizing yields the transformation of $\Psi_{0}$ (equation \ref{initial})
into $\Psi_{1}$ (equation \ref{fin}).
\end{itemize}

In hindsight, there is a shortcut. For each $\left\vert \mathbf{k}%
\right\rangle _{K}$,\ we perform function evaluation not only for the values
of $x$ required to identify the solution, also for the other values. I. e., we
perform function evaluation for each product $\left\vert \mathbf{k}%
\right\rangle _{K}\left(  \left\vert 0\right\rangle _{X}+\left\vert
1\right\rangle _{X}\right)  \left(  \left\vert 0\right\rangle _{V}-\left\vert
1\right\rangle _{V}\right)  $;\ junk histories (for that $\left\vert
\mathbf{k}\right\rangle _{K}$) do not harm, the important thing is performing
function evaluation for the values of $x$ required to identify the solution.
This yields the transformation of $\Psi_{0}$ into $\Psi_{1}$. Conversely, by
simply inspecting the form of $\Psi_{0}$ in equation (\ref{initial}), one can
see that each $\left\vert \mathbf{k}\right\rangle _{K}\left(  \left\vert
0\right\rangle _{X}+\left\vert 1\right\rangle _{X}\right)  \left(  \left\vert
0\right\rangle _{V}-\left\vert 1\right\rangle _{V}\right)  $ is the initial
state of a bunch of histories as from the shortcut. This shows that quantum
parallel computation is the sum of the histories of a classical algorithm
that, given the advanced information, computes the missing information
required to identify the solution of the problem. The final rotation of the
basis of register $X$\ serves to make the information about the oracle's
choice -- propagated to $X$ with function evaluation --\ readable.

Summing up, Deutsch's algorithm can be represented as a sum of the histories
of the related advanced information classical algorithm, with histories phases
and final rotation of the $X$\ basis that reconstruct the quantum algorithm.
Mutual causality thus explains the structure of the quantum algorithm.

Let us try and build the quantum algorithm out of the advanced information
classical algorithm, using no a priori knowledge of the quantum algorithm.
This requires finding a criteria for setting histories initial phases and
final rotation of the $X$\ basis.

Let us make a simplification: we still use a priori knowledge of the quantum
algorithm to set the amplitudes of the initial superposition in $X$ equal to
one another. Instead, we take the generic initial state of $V$: $\alpha\left(
\left\vert 0\right\rangle _{V}+\left\vert 1\right\rangle _{V}\right)
+\beta\left(  \left\vert 0\right\rangle _{V}-\left\vert 1\right\rangle
_{V}\right)  $. The initial amplitude of histories starting with $V$ in
$\left\vert 0\right\rangle _{V}$ is thus $\alpha+\beta$, with $V$ in
$\left\vert 1\right\rangle _{V}$ is $\alpha-\beta$. Under $\alpha$, the
computation performed by the advanced information classical algorithm gets
lost in the quantum translation, since the overall factorizable initial state
is transformed into itself. Under $\beta$, the transfer of information from
classical to quantum algorithm is maximum and the entanglement between
registers $K$\ and\ $X$\ is maximized (it is also maximal in the present
case). Maximizing the entanglement generated by function evaluation, yields
the function evaluation stage of Deutsch's algorithm. One can readily check
that this holds also if we set the amplitudes in the initial state of register
$X$\ free.

As for the rotation of the $X$\ basis, let us first discuss the form of the
state after function evaluation $\Psi_{1}$ (equation \ref{fin}). With
$\alpha=0$ and $\beta=1$ function evaluation has created a maximal
entanglement between registers $K$\ and $X$. Two orthogonal states of $K$,
$\left\vert 00\right\rangle _{K}-\left\vert 11\right\rangle _{K}$ and
$\left\vert 01\right\rangle _{K}-\left\vert 10\right\rangle _{K}$ are
correlated with two orthogonal states of $X$. This means that register
$X$\ contains the information that discriminates between $\left\vert
00\right\rangle _{K}-\left\vert 11\right\rangle _{K}$ and $\left\vert
01\right\rangle _{K}-\left\vert 10\right\rangle _{K}$ (or between
$\operatorname{e}^{i\delta_{00}}\left\vert 00\right\rangle _{K}%
-\operatorname{e}^{i\delta_{11}}\left\vert 11\right\rangle _{K}$ and
$\operatorname{e}^{i\delta_{01}}\left\vert 01\right\rangle _{K}%
-\operatorname{e}^{i\delta_{10}}\left\vert 10\right\rangle _{K}$ if the
superposition in $K$ is incoherent), namely between constant and balanced
functions. Therefore we should rotate the basis of $X$ in such a way that this
information becomes readable: $\left\vert 0\right\rangle _{X}+\left\vert
1\right\rangle _{X}$\ should go into $\left\vert 0\right\rangle _{X}$, etc. We
can define this rotation (independently of our a priori knowledge of the
quantum algorithm) by setting the requirement that the entanglement between
$K$ and $X$ (generated by function evaluation) becomes correlation between the
outcomes of measuring $\left[  K\right]  $ and $\left[  X\right]  $.

Summing up, in the case of Deutsch's algorithm, the advanced information
classical algorithm defines the quantum algorithm provided that history phases
maximize the entanglement generated by function evaluation and final rotation
of the $X$\ basis transforms this entanglement into correlation between
measurement outcomes.

The fact that function evaluation without other operations generates the
desired entanglement between character of the function and solution is not a
necessity of course. It is Deutsch's problem that is designed around this
fact. The problem could be different, for example distinguishing between
$f_{00}$ and $f_{01}$ on the one side and $f_{10}$ and $f_{11}$ on the other.
The algorithm should correspondingly be changed by left multiplying Hadamard
on $X$\ by the appropriate permutation of the basis vectors of $X$. In the
following, we will see that the problems addressed by the quantum algorithms
are all designed to exploit the entanglement generated by function evaluation
with the slightest use of other operations.

\section{Deutsch\&Jozsa's algorithm}

The set of functions is the constant and balanced functions $f_{\mathbf{k}%
}:\left\{  0,1\right\}  ^{n}\rightarrow\left\{  0,1\right\}  $; $\mathbf{k}%
\equiv k_{0},k_{1},...,k_{2^{n}-1}$ is both table and suffix -- table
(\ref{dj}) for $n=2$.
\begin{equation}%
\begin{tabular}
[c]{|c|c|c|c|c|c|c|c|c|}\hline
$x$ & $\,f_{0000}\left(  x\right)  $ & $f_{1111}\left(  x\right)  $ &
$f_{0011}\left(  x\right)  $ & $f_{1100}\left(  x\right)  $ & $f_{0101}\left(
x\right)  $ & $f_{1010}\left(  x\right)  $ & $f_{0110}\left(  x\right)  $ &
$f_{1001}\left(  x\right)  $\\\hline
00 & 0 & 1 & 0 & 1 & 0 & 1 & 0 & 1\\\hline
01 & 0 & 1 & 0 & 1 & 1 & 0 & 1 & 0\\\hline
10 & 0 & 1 & 1 & 0 & 0 & 1 & 1 & 0\\\hline
11 & 0 & 1 & 1 & 0 & 1 & 0 & 0 & 1\\\hline
\end{tabular}
\label{dj}%
\end{equation}

The oracle chooses a function. The problem is finding whether the function is
balanced or constant through function evaluation $\left[  7\right]  $.

The black box, given $\mathbf{k}$ and $x$, computes $f\left(  \mathbf{k}%
,x\right)  =f_{\mathbf{k}}(x)$. The $2^{n}$ qubit oracle's choice \ register
$K$\ (just a conceptual reference) contains $\mathbf{k}$. The algorithm is:
(0) prepare $K$ in the superposition of all $\mathbf{k}$, $X$ in the
superposition of all $x$, and $V$ in the antisymmetric state, (1)\ perform
function evaluation, which changes the content of $V$ from $v$ to $v\oplus
f\left(  \mathbf{k},x\right)  $, and (2) apply Hadamard to register $X$:%

\begin{align}
\Psi_{0}  &  =\frac{1}{8}\left(  \left\vert 0000\right\rangle _{K}+\left\vert
1111\right\rangle _{K}+\left\vert 0011\right\rangle _{K}+\left\vert
1100\right\rangle _{K}+...\right) \label{indj}\\
&  \left(  \left\vert 00\right\rangle _{X}+\left\vert 01\right\rangle
_{X}+\left\vert 10\right\rangle _{X}+\left\vert 11\right\rangle _{X}\right)
\left(  \left\vert 0\right\rangle _{V}-\left\vert 1\right\rangle _{V}\right)
.\nonumber
\end{align}

\begin{equation}
\Psi_{1}=\frac{1}{8}\left[
\begin{array}
[c]{c}%
(\left\vert 0000\right\rangle _{K}-\left\vert 1111\right\rangle _{K}%
)(\left\vert 00\right\rangle _{X}+\left\vert 01\right\rangle _{X}+\left\vert
10\right\rangle _{X}+\left\vert 11\right\rangle _{X})+\\
(\left\vert 0011\right\rangle _{K}-\left\vert 1100\right\rangle _{K}%
)(\left\vert 00\right\rangle _{X}+\left\vert 01\right\rangle _{X}-\left\vert
10\right\rangle _{X}-\left\vert 11\right\rangle _{X})+...
\end{array}
\right]  \left(  \left\vert 0\right\rangle _{V}-\left\vert 1\right\rangle
_{V}\right)  . \label{evaluation}%
\end{equation}

\begin{equation}
\Psi_{2}=\frac{1}{4}\left[  \left(  \left\vert 0000\right\rangle
_{K}-\left\vert 1111\right\rangle _{K}\right)  \left\vert 00\right\rangle
_{X}+(\left\vert 0011\right\rangle _{K}-\left\vert 1100\right\rangle
_{K})\left\vert 10\right\rangle _{X}+....\right]  \left(  \left\vert
0\right\rangle _{V}-\left\vert 1\right\rangle _{V}\right)  . \label{hdj}%
\end{equation}
Measuring $\left[  K\right]  $\ and $\left[  X\right]  $\ in (\ref{hdj})
determines the oracle's choice and the solution found by the second player:
all zeroes if constant, not so if balanced. Backdating the reduction on
$\mathbf{k}$\ gives the original algorithm.

We check the 50\% rule. There is advanced knowledge of any half $\mathbf{k}$.
We can distinguish between: (i) half tables that do not contain different
values of the function and (ii) those that do. In (i),\ the solution is
identified by computing an extra row, i. e. by performing one function
evaluation for any value of $x$\ outside the advanced information (if the
value of the function is still the same, the function is constant, otherwise
it is balanced). In (ii), we know already that the function is balanced and no
function evaluation is needed. Although some half $\mathbf{k}$\ require one
function evaluation and some none, the 50\% rule -- the fact that a quantum
algorithm requires the number of function evaluations of a classical algorithm
that knows in advance 50\% of the information about solution of the
problem\ -- is satisfied. In fact the rule excludes the half $\mathbf{k}%
$\ that already specify the solution and require no function evaluation,
because they contain 100\% of the information about the solution.

The function evaluation stage of the quantum algorithm (the transformation of
$\Psi_{0}$ into $\Psi_{1}$) is the sum of the histories of the advanced
information classical algorithm as from\ the "shortcut" highlighted for
Deutsch's algorithm.

We build the quantum algorithm out of the advanced information classical
algorithm. As for the choice of the history initial phases, this is justified
as in Deutsch's algorithm. As for the rotation of the $X$\ basis, we examine
the outcome of function evaluation, namely $\Psi_{1}$ (equation
\ref{evaluation}). There is maximal entanglement between $K$\ and
$X$.\ Orthogonal states of $K$, discriminating between constant and balanced
functions, are correlated with orthogonal states of $X$. The information
whether the function is constant or balanced has propagated to register $X$.
To read this information, we rotate the $X$ basis in such a way that
$(\left\vert 0000\right\rangle _{K}-\left\vert 1111\right\rangle
_{K})(\left\vert 00\right\rangle _{X}+\left\vert 01\right\rangle
_{X}+\left\vert 10\right\rangle _{X}+\left\vert 11\right\rangle _{X})$\ goes
into $(\left\vert 0000\right\rangle _{K}-\left\vert 1111\right\rangle
_{K})\left\vert 00\right\rangle _{X}$, etc. This rotation of the basis of $X$
is such that the entanglement between $K$ and $X$ becomes correlation between
the outcomes of measuring $\left[  K\right]  $ and $\left[  X\right]  $. This
is a constructive definition of Hadamard on $X$. This completes the derivation
of the quantum algorithm from the advanced information classical algorithm.

\section{Bernstein\&Vazirani's algorithm}

The analysis of the previous section also applies to Bernstein\&Vazirani's
algorithm $\left[  1\right]  $, which is a restriction of Deutsch\&Jozsa's
algorithm. The set of the constant and balanced functions is restricted to a
proper subset thereof, namely to the functions such that $f_{\mathbf{k}%
}(x)=a\cdot x$ , with $a\cdot x=\left(  \sum_{i\in\left\{  0,1\right\}  ^{n}%
}a_{i}x_{i}\right)  \operatorname{mod}2$. The problem is to find the "hidden
string" $a$. The algorithm is Deutsch\&Jozsa's algorithm with the
superposition hosted in register $K$ reduced to the new set of functions.
Measuring $\left[  K\right]  $\ and $\left[  X\right]  $\ at the end of the
algorithm yields a value of $\mathbf{k}$\ and the corresponding value of $a$.
The discussion is the same as in the former section.

\section{Simon's algorithm}

The set of functions is $f_{\mathbf{k}}:\left\{  0,1\right\}  ^{n}%
\rightarrow\left\{  0,1\right\}  ^{n-1}$, such that $f_{\mathbf{k}}\left(
x\right)  =f_{\mathbf{k}}\left(  y\right)  $ iff $x=y$\ or $x=y\oplus
\mathbf{h}^{\left(  \mathbf{k}\right)  }$; $\mathbf{h}^{\left(  \mathbf{k}%
\right)  }\mathbf{\equiv~}h_{0}^{\left(  \mathbf{k}\right)  },h_{1}^{\left(
\mathbf{k}\right)  },...,h_{n-1}^{\left(  \mathbf{k}\right)  }$ (depending on
$\mathbf{k}$) belongs to $\left\{  0,1\right\}  ^{n}$ excluding the all zeroes
string; $\oplus$\ denotes bitwise $\operatorname{mod}2$ addition. Table
(\ref{periodic}) gives the set for $n=2$; $\mathbf{k}$ is both table and
suffix. Since $\mathbf{h}^{\left(  \mathbf{k}\right)  }\oplus\mathbf{h}%
^{\left(  \mathbf{k}\right)  }=0$, each value of the function appears exactly
twice in the table, thus 50\% of the rows plus one surely identify
$\mathbf{h}^{\left(  \mathbf{k}\right)  }$.
\begin{equation}%
\begin{tabular}
[c]{|c|c|c|c|c|c|c|}\hline
& $\mathbf{h}^{\left(  0011\right)  }=01$ & $\mathbf{h}^{\left(  1100\right)
}=01$ & $\mathbf{h}^{\left(  0101\right)  }=10$ & $\mathbf{h}^{\left(
1010\right)  }=10$ & $\mathbf{h}^{\left(  0110\right)  }=11$ & $\mathbf{h}%
^{\left(  1001\right)  }=11$\\\hline
$x$ & $f_{0011}\left(  x\right)  $ & $f_{1100}\left(  x\right)  $ &
$f_{0101}\left(  x\right)  $ & $f_{1010}\left(  x\right)  $ & $f_{0110}\left(
x\right)  $ & $f_{1001}\left(  x\right)  $\\\hline
00 & 0 & 1 & 0 & 1 & 0 & 1\\\hline
01 & 0 & 1 & 1 & 0 & 1 & 0\\\hline
10 & 1 & 0 & 0 & 1 & 1 & 0\\\hline
11 & 1 & 0 & 1 & 0 & 0 & 1\\\hline
\end{tabular}
\ \label{periodic}%
\end{equation}

The oracle chooses a function. The problem is finding the value of
$\mathbf{h}^{\left(  \mathbf{k}\right)  }$, "hidden" in the $f_{\mathbf{k}%
}\left(  x\right)  $ chosen by the oracle, through function evaluation
$\left[  12\right]  $. The black box, given $\mathbf{k}$ and $x$, computes
$f\left(  \mathbf{k},x\right)  =f_{\mathbf{k}}(x)$. The oracle's choice
\ register $K$\ is $2^{n}\left(  n-1\right)  $ qubit. The algorithm is: (0)
prepare $K$ in the superposition of all $\mathbf{k}$, $X$ in the superposition
of all $x$, and $V$ in the all zeroes string $\left\vert \mathbf{0}%
\right\rangle _{V}$, (1)\ perform function evaluation, which changes the
content of $V$ from $\mathbf{v}$ to $\mathbf{v}\oplus f\left(  \mathbf{k}%
,x\right)  $, where $\oplus$ denotes bitwise $\operatorname{mod}2$ addition,
and (2) apply Hadamard to $X$:%

\begin{equation}
\Psi_{0}=\frac{1}{2\sqrt{6}}\left(  \left\vert 0011\right\rangle
_{K}+\left\vert 1100\right\rangle _{K}+\left\vert 0101\right\rangle
_{K}+\left\vert 1010\right\rangle _{K}+...\right)  \left(  \left\vert
00\right\rangle _{X}+\left\vert 01\right\rangle _{X}+\left\vert
10\right\rangle _{X}+\left\vert 11\right\rangle _{X}\right)  \left\vert
0\right\rangle _{V}. \label{insimon}%
\end{equation}

\begin{equation}
\Psi_{1}=\frac{1}{2\sqrt{6}}\left[
\begin{array}
[c]{c}%
(\left\vert 0011\right\rangle _{K}+\left\vert 1100\right\rangle _{K})\left[
(\left\vert 00\right\rangle _{X}+\left\vert 01\right\rangle _{X})\left\vert
0\right\rangle _{V}+(\left\vert 10\right\rangle _{X}+\left\vert
11\right\rangle _{X})\left\vert 1\right\rangle _{V}\right]  +\\
(\left\vert 0101\right\rangle _{K}+\left\vert 1010\right\rangle _{K})\left[
(\left\vert 00\right\rangle _{X}+\left\vert 10\right\rangle _{X})\left\vert
0\right\rangle _{V}+(\left\vert 01\right\rangle _{X}+\left\vert
11\right\rangle _{X})\left\vert 1\right\rangle _{V}\right]  +...
\end{array}
\right]  . \label{second}%
\end{equation}

\begin{equation}
\Psi_{2}=\frac{1}{2\sqrt{6}}\left[
\begin{array}
[c]{c}%
(\left\vert 0011\right\rangle _{K}+\left\vert 1100\right\rangle _{K})\left[
(\left\vert 00\right\rangle _{X}+\left\vert 10\right\rangle _{X})\left\vert
0\right\rangle _{V}+(\left\vert 00\right\rangle _{X}-\left\vert
10\right\rangle _{X})\left\vert 1\right\rangle _{V}\right]  +\\
(\left\vert 0101\right\rangle _{K}+\left\vert 1010\right\rangle _{K})\left[
(\left\vert 00\right\rangle _{X}+\left\vert 01\right\rangle _{X})\left\vert
0\right\rangle _{V}+(\left\vert 00\right\rangle _{X}-\left\vert
01\right\rangle _{X})\left\vert 1\right\rangle _{V}\right]  +...
\end{array}
\right]  . \label{hsimon}%
\end{equation}

In $\Psi_{2}$, for each value of the content of $K$ and no matter the content
of $V$, register $X$ hosts even weighted\ superpositions of the $2^{n-1}$
strings $\mathbf{s}_{j}^{\left(  \mathbf{k}\right)  }$ orthogonal to
$\mathbf{h}^{\left(  \mathbf{k}\right)  }$. By measuring $\left[  K\right]
$\ and $\left[  X\right]  $ we obtain at random the oracle's choice
$\mathbf{k}$ and one of the $\mathbf{s}_{j}^{\left(  \mathbf{k}\right)  }$.

We leave $K$ in its after-measurement state, thus fixing $\mathbf{k}$, and
iterate the right part of the algorithm (preparation of registers $X$\ and
$V$, function evaluation, and measurement of $\left[  X\right]  $) until
obtaining $n-1$ different $\mathbf{s}_{j}^{\left(  \mathbf{k}\right)  }$,
which allows to find $\mathbf{h}^{\left(  \mathbf{k}\right)  }$ by solving the
system of $n-1$ $\operatorname{mod}2$ linear equations.

We check the 50\% rule. We focus on the quantum part of Simon's problem,
namely on the problem of generating at random a string $\mathbf{s}%
_{j}^{\left(  \mathbf{k}\right)  }$ orthogonal to $\mathbf{h}^{\left(
\mathbf{k}\right)  }$. Any $\mathbf{s}_{j}^{\left(  \mathbf{k}\right)  }$ is a
"solution" of this problem. This formulation of Simon's problem and the usual
formulation (finding $\mathbf{h}^{\left(  \mathbf{k}\right)  }$) are
equivalent as far as an exponential speed up in the former implies an
exponential speed up in the latter and vice-versa. The information acquired by
measuring $\left[  K\right]  $ and $\left[  X\right]  $ in (\ref{hsimon})\ is
the information in $K$ (the content of $X$ -- the string $\mathbf{s}%
_{j}^{\left(  \mathbf{k}\right)  }$ -- is a function of the content of $K$).
Measuring $\left[  K\right]  $ induces state reduction on $\mathbf{k}$. There
is advanced knowledge of anyone of the two halves of $\mathbf{k}$. If the half
table does not contain a same value of the function twice, the solution is
identified by performing one function evaluation for any value of $x$\ outside
the advanced information. The new value of the function is necessarily a value
already present in the advanced information, which identifies $\mathbf{h}%
^{\left(  \mathbf{k}\right)  }$ thus all the $\mathbf{s}_{j}^{\left(
\mathbf{k}\right)  }$. If the half table contains a same value of the function
twice, this already identifies the solution and no function evaluation is
needed. This verifies the 50\% rule.

One can see that the same discussion applies to the Generalized Simon's
algorithm, thus to the Hidden Subgroup Algorithms, like finding the period of
a function (the quantum part of Shor's factorization algorithm), etc.

The function evaluation stage of the quantum algorithm is the sum of the
histories of the advanced information classical algorithm -- as from the
shortcut highlighted for Deutsch's algorithm.

As for building the quantum algorithm out of the advanced information
classical algorithm, things are more difficult. In fact Simon's algorithm is
not optimal under the criteria of maximizing entanglement. We show this for
$n=2$. We still choose even amplitudes for the initial superposition in $X$,
while letting $V$ in the generic initial state $\alpha\left\vert
0\right\rangle _{V}+\beta\left\vert 1\right\rangle _{V}$. As readily checked,
$\alpha=-\beta$ maximizes the entanglement between $K$\ and $X$ after function
evaluation. Hadamard on $X$ transforms this entanglement into correlation
between the outcomes of measuring $\left[  K\right]  $ and $\left[  X\right]
$.

With $V$\ prepared in the antisymmetric state, state (\ref{hsimon}) becomes:%

\[
\frac{1}{2\sqrt{3}}\left[  (\left\vert 0011\right\rangle _{K}+\left\vert
1100\right\rangle _{K})\left\vert 10\right\rangle _{X}+(\left\vert
0101\right\rangle _{K}+\left\vert 1010\right\rangle _{K})\left\vert
01\right\rangle _{X}+...\right]  \left(  \left\vert 0\right\rangle
_{V}-\left\vert 1\right\rangle _{V}\right)  .
\]
Swapping the basis vectors $\left\vert 01\right\rangle _{X}$ and $\left\vert
10\right\rangle _{X}$ yields the desired correlation between $\mathbf{k}$ and
$\mathbf{h}^{\left(  \mathbf{k}\right)  }$. In terms of number of oracle's
queries, the optimal algorithm is more efficient than the known algorithm. In
fact it yields the hidden string $\mathbf{h}^{\left(  \mathbf{k}\right)  }$
(rather than a $\mathbf{s}_{j}^{\left(  \mathbf{k}\right)  }$) with a single
oracle's query, as expected since just one function evaluation identifies
$\mathbf{h}^{\left(  \mathbf{k}\right)  }$. Simon's algorithm is suboptimal in
terms of number of oracle's queries, however it scales up to any value of $n$.
It is not clear that this is the case for the "optimal" algorithm.

\section{Grover's algorithm}

See section 2 for $n=2$. We check the 50\% rule for $n=2$. There is advanced
knowledge of anyone of the two bits in the outcome of measuring $\left[
K\right]  $. The quantum algorithm requires the number of function evaluations
of a classical algorithm that knows in advance either $k_{0}$ or $k_{1}$. To
identify the missing bit, this algorithm should perform one function
evaluation, as readily checked. This verifies the 50\% rule. As for $n>2$, a
classical algorithm that knows in advance 50\% of the $n$\ bits of the data
base location, to identify the $n/2$\ missing bits should perform $O\left(
2^{n/2}\right)  $ function evaluations, against the $O\left(  2^{n}\right)  $
of a classical algorithm without advanced information. This verifies the 50\%
rule for $n>2$.

As for representing the quantum algorithm as the sum of the histories of the
advanced information classical algorithm, still for $n=2$, the procedure is
the same as in the former algorithms.

We build the quantum algorithm out of the advanced information classical
algorithm -- still for $n=2$ to start with. As for the history initial phases,
their choice is justified as in Deutsch's algorithm. As for the final rotation
of the $X$\ basis, we examine the outcome of function evaluation, i. e.
$\Psi_{1}$ (equation \ref{secondstage}). Registers $K$ and $X$ are maximally
entangled, orthogonal states of $K$, each corresponding to a value of
$\mathbf{k}$, are correlated with orthogonal states of $X$, which means that
the value of $\mathbf{k}$\ has propagated to register $X$. To read this value,
i. e. to transform entanglement between $K$ and $X$ into correlation between
the outputs of measuring $\left[  K\right]  $ and $\left[  X\right]  $, we
should rotate the $X$ basis in such a way that $-\left\vert 00\right\rangle
_{X}+\left\vert 01\right\rangle _{X}+\left\vert 10\right\rangle _{X}%
+\left\vert 11\right\rangle _{X}$ (correlated with $\left\vert 00\right\rangle
_{K}$) goes into $\left\vert 00\right\rangle _{X}$, etc. This is a
constructive definition of the transformation $U$.

Generalizing to $n>2$ is straightforward. Given the advanced knowledge of
$n/2$ bits, in order to compute the missing $n/2$ bits we should perform
function evaluation and rotation of the basis of $X$ an\ $\operatorname{O}%
\left(  2^{\frac{n}{2}}\right)  $ times. Each time we obtain the superposition
of an unentangled state of the form (\ref{preparation}) and a maximally
entangled state of the form (\ref{final}). At each successive iteration, the
amplitude of the latter state is amplified at the expense of the amplitude of
the former, until it becomes about $1$ in $\operatorname{O}\left(  2^{\frac
{n}{2}}\right)  $ iterations. The final measurement of $\left[  K\right]  $
induces state reduction on $\mathbf{k}$. Mutual causality is between any two
halves of $\mathbf{k}$.

\section{Engineering quantum algorithms}

The 50\% rule and the sum of the histories representation of a quantum
algorithm make up a tool for the search of quantum speed ups. The rule, by
comparing two classical algorithms -- with and without advanced information --
allows to identify the problems liable of being solved with a quantum speed up
in terms of number of oracle's queries. Once identified such a problem, the
sum of the histories of the advanced information classical algorithm -- with
histories degrees of freedom set to maximize entanglement/correlation --
should yield this speed up. This can be a speed up in the overall number of
operations, like in Deutsch's, Deutsch\&Jozsa's, Bernstein\&Vazirani's, and
Grover's algorithms. On the contrary, the speed up in terms of number of
oracle's queries can in principle be frustrated by the cost of other
operations, like e. g. setting the histories degrees of freedom. In Simon's
algorithm, one gets an overall speed up with a suboptimal number of queries.

We exemplify the application of this tool by building from scratch a pair of
simple quantum algorithms.

The first is a variation of Deutsch's algorithm. The oracle chooses a function
out of the set $f_{\mathbf{k}}:\left\{  0,1\right\}  \rightarrow\left\{
00,01,10\right\}  $ -- table (\ref{minute}).%

\begin{equation}%
\begin{tabular}
[c]{|c|c|c|c|c|c|c|c|c|c|}\hline
$x$ & $f_{0000}$ & $f_{0001}$ & $f_{0100}$ & $f_{0101}$ & $f_{0010}$ &
$f_{1000}$ & $f_{1001}$ & $f_{0110}$ & $f_{1010}$\\\hline
0 & 00 & 00 & 01 & 01 & 00 & 10 & 10 & 01 & 10\\\hline
1 & 00 & 01 & 00 & 01 & 10 & 00 & 01 & 10 & 10\\\hline
\end{tabular}
\ \ \label{minute}%
\end{equation}

The problem is whether the $\operatorname{mod}2$ addition of the four bits of
$\mathbf{k}$ is 0 or 1. Given any half table (the value of the function for a
single value of $x$), the advanced information classical algorithm, in order
to find the solution, should perform function evaluation for the other value
of $x$.

We derive the quantum algorithm out of the advanced information classical
algorithm. We need a 4 qubits oracle's choice register $K$, a 1 qubit oracle's
query register $X$, and a 2 qubits register $V$\ for the result of function
evaluation (bitwise $\operatorname{mod}2$ added to the former content of $V$
for logical reversibility). Setting histories degrees of freedom to maximize
entanglement/correlation yields the algorithm: (0) prepare $K$ in the
superposition of all $\mathbf{k}$, $X$ in the superposition of all $x$, each
qubit of $V$ in the antisymmetric state, (1)\ perform function evaluation, and
(2) apply Hadamard on $X$. The final measurement of $\left[  K\right]  $ and
$\left[  X\right]  $ yields the function chosen by the oracle in $K$ and the
solution in $X$.

We provide another example. Let the set of functions be the $4!$\ functions
$f_{\mathbf{k}}:\left\{  0,1\right\}  ^{2}\rightarrow\left\{  0,1\right\}
^{2}$ such that the sequence of function values is a permutation of the values
of the argument -- table (\ref{perm}).%

\begin{equation}%
\begin{tabular}
[c]{|c|c|c|c|}\hline
$x$ & $f_{00011011}$ & $f_{00011110}$ & $...$\\\hline
00 & 00 & 00 & ...\\\hline
01 & 01 & 01 & ...\\\hline
10 & 10 & 11 & ...\\\hline
11 & 11 & 10 & ...\\\hline
\end{tabular}
\ \ \ \label{perm}%
\end{equation}
A classical algorithm with the advanced information -- 50\% of the rows of
$f_{\mathbf{k}}$\ -- identifies the function with just one function
evaluation. Without, with three function evaluations. There is room for a
speed up in terms of number of oracle's queries. We build a quantum algorithm
over this possibility. Register $K$\ is $8$ qubits, registers $X$ is $2$
qubits, and register $V$ is $2$\ qubits. The preparation is%

\begin{equation}
\frac{1}{8\sqrt{6}}\left(  \left\vert 00011011\right\rangle _{K}+\left\vert
00011110\right\rangle _{K}+...\right)  \left(  \left\vert 00\right\rangle
_{X}+\left\vert 01\right\rangle _{X}+\left\vert 10\right\rangle _{X}%
+\left\vert 11\right\rangle _{X}\right)  \left(  \left\vert 0\right\rangle
_{V_{0}}-\left\vert 1\right\rangle _{V_{0}}\right)  \left(  \left\vert
0\right\rangle _{V_{1}}-\left\vert 1\right\rangle _{V_{1}}\right)  .
\label{inperm}%
\end{equation}
Performing just one function evaluation and then Hadamard on $X$, yields an
entangled state where three orthogonal states of $K$ (each a superposition of
8 values of $\mathbf{k}$, corresponding to a partition of the set of 24
functions) are correlated with, respectively, $\left\vert 01\right\rangle
_{X},~\left\vert 10\right\rangle _{X},~$and $\left\vert 11\right\rangle _{X}$.
Measuring $\left[  X\right]  $ in (\ref{inperm}) tells which of the three
partitions the function belongs to. In the case of a classical algorithm, this
requires three function evaluations. There is a quantum speed up.

With the 50\% rule, one can figure out any number of these speed ups. The
problem is of course finding speed ups of practical interest. Anyway, this
exemplifies how the 50\% rule and the sum of the histories yield a playground
for searching new quantum speed ups and developing quantum algorithm engineering.

\section{Conclusions}

Summing up, the points in favour of the 50\% rule are: (i) the rule is self
evident, it is an immediate consequence of postulating the symmetry of mutual
causality (like between two polarizations measured in an entangled
polarization state)\ in a more complete representation of the quantum
algorithms, (ii)\ the rule is verified by the main quantum algorithms, and
(iii)\ the rule explains the structure of these algorithms -- which are in
fact representable as a sum of the histories of the respective advanced
information classical algorithm.

The 50\% rule and the sum of the histories picture make up a tool for the
search of quantum speed ups, as discussed in the previous section. This should
be interesting in a situation where finding a speed up was, so to speak, a
matter of art and no further speed ups of practical interest have been
discovered since 1996. Using this tool for a systematic search of the speed
ups should also foster the quantum algorithms engineering outlined in this paper.

This work should provide a better understanding of the foundations of quantum
information. Even if quantum information is more and more becoming an
experimental science and changing to quantum control, it should not forget its
foundational problems, namely scarcity of quantum speed ups and decoherence.
In principle the two leads, theoretical and experimental, should advance together.

Because of its peculiar character, the present explanation of the speed up
could have an epistemological and interdisciplinary interest. It questions the
mechanistic vision of evolutions where causality propagates locally through an
instantaneous present (the $dt$ of Schr\"{o}dinger equation). Quantum
algorithms are partly driven by their future outcome and, in some sense,
"exist" (host mutual causality)\ in an extended present comprising
preparation, unitary evolution, and measurement. Under the perspective of the
50\% rule, quantum computation turns out to be the first formalized example of
teleological evolution. This might shed light on quantum and teleological
explanations of organic behavior.

\subsubsection{Acknowledgements}

The author thanks for useful comments/suggestions his wife Ferdinanda Pavia
and: Scott Aaronson, Pablo Arrighi, Vint Cerf, Artur Ekert, David Finkelstein,
Shlomit Finkelstein, Hartmut Neven, and Charles Stromeyer.

\subsubsection*{Bibliography}

$1.$ \ \ Bernstein, E. \& Vazirani, U. Proceedings of the 25th annual ACM
Symposium on the Theory of Computing, 11 (1993)

$2.$ \ \ Castagnoli, G. \& Finkelstein, D.: Theory of the quantum speed up.
Proc. Roy. Soc. Lond. A \textbf{457}, 1799 . arXiv:quant-ph/0010081 v1 (2001)

$3.$ \ \ Castagnoli, G.: The mechanism of quantum computation. Int. J. Theor.
Phys.,\textit{\ }vol. \textbf{47}, number 8, 2181 (2008)

$4.$\ \ \ Castagnoli, G.: The quantum speed up as advanced cognition of the
solution. Int. J. Theor. Phys., vol. \textbf{48}, issue 3, 857 (2009)

$5.$\ \ \ Castagnoli, G.: The 50\% advanced information rule of the quantum
algorithms. To be published in Int. J. Theor. Phys. (2009)

$6.$ \ \ Deutsch, D.: Quantum theory, the Church-Turing principle, and the
universal quantum computer. Proc. Roy. Soc. (Lond.) A, \textbf{400}, 97 (1985)

$7.$ \ \ Deutsch, D. \& Jozsa, R.: Rapid solution of problems by quantum
computation. Proc. Roy. Soc. (Lond.) A, \textbf{439}, 553 (1992)

$8.$\ \ \ Dolev, S. \&\ Elitzur, A. C.: Non-Sequential Behavior of the Wave
Function. arXiv:quant-ph/0102109 v1 (2005)

$9.$ \ \ Gross, D., Flammia, S. T. \& Eisert, J. Phys. Rev. Lett. volume 102,
issue 19 (2009)

$10.$ Grover, L. K.: A fast quantum mechanical algorithm for database search.
Proc. 28th Ann. ACM Symp. Theory of Computing (1996)

$11.$\ Grover, L. K.: From Schr\"{o}dinger equation to quantum search
algorithm. arXiv: quant-ph/0109116 (2001)

$12.$ Simon, D.: On the power of quantum computation. Proc. 35th Ann. Symp. on
Foundations of Comp. Sci.,\textit{\ }116 (1994)

\end{document}